\documentclass[sigconf]{acmart}

\AtBeginDocument{%
  }

\setcopyright{none}

\acmConference[Human-centered Explainable AI Workshop (HCXAI) @ CHI 2026]{}{April 13, 2016}{Barcelona, Spain}

\begin{document}


\title[Explainable AI for Blind and Low-Vision Users]{Explainable AI for Blind and Low-Vision Users: Navigating Trust, Modality, and Interpretability in the Agentic Era}


\author{Abu Noman Md Sakib}
\email{abunomanmd.sakib@utsa.edu}
\orcid{0000-0002-0761-035X}
\affiliation{%
  \institution{University of Texas at San Antonio}
  \city{San Antonio}
  \state{Texas}
  \country{USA}
}

\author{Protik Dey}
\email{protik.dey@utsa.edu}
\orcid{0009-0006-4089-5111}
\affiliation{%
  \institution{University of Texas at San Antonio}
  \city{San Antonio}
  \state{Texas}
  \country{USA}
}

\author{Zijie Zhang}
\email{zijie.zhang@utsa.edu}
\orcid{0000-0003-1254-098X}
\affiliation{%
  \institution{University of Texas at San Antonio}
  \city{San Antonio}
  \state{Texas}
  \country{USA}
}

\author{Taslima Akter}
\email{taslima.akter@utsa.edu}
\orcid{0000-0002-0686-1448}
\affiliation{%
  \institution{University of Texas at San Antonio}
  \city{San Antonio}
  \state{Texas}
  \country{USA}
}

\renewcommand{\shortauthors}{A.N.M. Sakib et al.}

\begin{abstract}
Explainable Artificial Intelligence (XAI) is critical for ensuring trust and accountability, yet its development remains predominantly visual. For blind and low-vision (BLV) users, the lack of accessible explanations creates a fundamental barrier to the independent use of AI-driven assistive technologies. This problem intensifies as AI systems shift from single-query tools into autonomous agents that take multi-step actions and make consequential decisions across extended task horizons, where a single undetected error can propagate irreversibly before any feedback is available. This paper investigates the unique XAI requirements of the BLV community through a comprehensive analysis of user interviews and contemporary research. By examining usage patterns across environmental perception and decision support, we identify a significant modality gap. Empirical evidence suggests that while BLV users highly value conversational explanations, they frequently experience ``self-blame'' for AI failures. The paper concludes with a research agenda for accessible Explainable AI in agentic systems, advocating for multimodal interfaces, blame-aware explanation design, and participatory development.
\end{abstract}




\maketitle

\section{Introduction}
Integrating Artificial Intelligence (AI) into daily life offers incredible autonomy for the blind and low-vision (BLV) community \cite{misfitAI, humanError}. With the introduction of vision models and large language models (LLMs), AI systems can now describe scenes, recognize objects, and interpret complex environments in real time. For many BLV individuals, these tools are no longer just convenient. They have become essential aids that change how people navigate the physical world \cite{MAIDR}. However, these systems are rapidly evolving from simple description tools into autonomous agents that plan multi-step task sequences, invoke external tools, and take actions with real-world consequences. Because their inner workings remain a ``black box,'' this shift creates an especially acute challenge for the BLV community \cite{teachableAI}. A sighted user who sees an agent draft a message or confirm a purchase can interrupt if something looks wrong. A BLV user has no equivalent monitoring channel. An error in Step~2 of a 10-step task can propagate silently through all subsequent steps and result in an irreversible outcome before any feedback is received. An AI system's output is often the primary source of environmental truth for a BLV user. Sometimes, it is the only source. Without accessible ways to inspect how the agent made its decisions, users cannot detect hallucinations, erroneous tool invocations, or cascading misinterpretations \cite{pieceIt}. The current approach to Explainable AI (XAI) makes this problem worse. The Human-Centered Explainable AI (HCXAI) movement tries to look beyond math and understand how people actually use explanations \cite{HCXAI}. Yet, mainstream XAI methods like SHAP \cite{lundberg2017unified} and GradCAM \cite{selvaraju2017grad} still rely heavily on visuals. By using color-coded heatmaps and visual bounding boxes, traditional XAI shuts out the exact users who rely on AI the most for sensory translation.

This paper addresses that gap by exploring accessible XAI and user trust within the BLV community. We look at how trust changes depending on the stakes of the situation. We also highlight a common ``Self-Blame Bias.'' This occurs when BLV users blame themselves for taking a bad photo rather than realizing the AI made a mistake. We then evaluate which types of explanations work best for verifying information and supporting decisions, comparing formats like system rules against conversational interactions. We argue that real digital inclusion for the BLV community requires legitimacy across the entire AI pipeline. This means everything from using representative training data to building clear, non-visual ways for users to understand AI decisions.

\section{Related Work}
Recent work in HCXAI highlights that explanations must be tailored to the people actually using them \cite{HCXAI, HCXAIGynae, ehsan2023human}. However, most mainstream methods for computer vision still rely on visual cues like heatmaps, which completely exclude BLV users \cite{treemap}. To fix this, accessible XAI focuses on translating model logic into non-visual formats like text or conversation. This is especially important because BLV users face unique challenges with trust. As AI moves from simple object recognition to complex, multi-step autonomous agents, this overreliance creates serious epistemological vulnerability \cite{AIreliance, AIRelianceSurvey}. If a system confidently hallucinates early in a task, the user has no easy way to catch the error \cite{hallucinationHCXAI}. Researchers have suggested using conversational AI to introduce ``cognitive forcing,'' encouraging users to ask questions and double-check outputs rather than passively accepting them \cite{conversationalAI, relationalAI}. Some studies indicate that BLV users develop complex, often flawed, mental models of Generative AI, sometimes viewing these tools as authoritative rather than probabilistic systems prone to hallucination \cite{kingofknowledge}. Furthermore, trust in these systems is highly contextual. Xinru et al. demonstrated in their framework that users' acceptance of technical failures is not binary but contingent on the stakes of the task and the social environment \cite{EverydayUncertainty}. Accessible XAI must support explorable interactions that allow users to interrogate these stakes \cite{explorablexai}. We view all of these challenges which argues that an AI system is only valid if it centers the people impacted by its decisions. For the BLV community, this means building legitimacy directly into the pipeline, starting with training data that actually reflects how they take photos, all the way to providing transparent, non-visual ways to verify what the AI is doing.

The emergence of LLM-based agents introduces explainability demands that go beyond explaining a single model prediction. Agentic systems orchestrate chains of reasoning steps, invoke external tools, and produce execution traces that unfold over time \cite{hallucinationHCXAI}. The relevant unit of explanation is no longer a feature attribution score but a process trace: which sub-goals were set, which tools were called, and where the agent chose to act rather than verify. The proxy verification strategies BLV users rely on today (barcode cross-checking, multi-shot photography, human fallback) are workarounds for the absence of accessible process transparency. These workarounds do not scale to agentic pipelines. A user cannot take three photos of an agent's multi-step scheduling decision. This makes the BLV community a critical test case for agentic XAI design.

\section{Methodology}
We conducted semi-structured interviews with six BLV participants so far with prior experience using GenAI tools, including general-purpose systems such as ChatGPT, Google Gemini, Meta AI, and Microsoft Copilot, as well as GenAI-powered assistive technologies such as Be My AI, Seeing AI, and Aira. Recruitment was conducted through the National Federation of the Blind, with a small number of participants recruited via snowball sampling. Interested individuals completed an online recruitment form, which also served as a screening form to assess eligibility. Eligibility criteria included: (1)~residing in the United States, (2)~age 18 years or older, (3)~identify as blind or low vision, (4)~primarily use a screen reader for accessing digital content (instead of magnification), (5)~speak English, and (6)~have experience of using GenAI tools. One researcher contacted each eligible participant via email to schedule the interviews.

Most of the participants are moderate users of GenAI tools, with two participants reported using them frequently, and one participant mentioned having comparatively less experience. All participants use GenAI-powered assistive technologies such as Be My AI, Seeing AI, Aira, etc. To access these GenAI tools, all participants reported using screen readers like JAWS, NVDA, and VoiceOver. Out of the six participants, five were females (83.33\%) and one participant was male (16.67\%). One participant was in the 18-29 age group, three were in the 40-49 range, one reported being between 50 to 59 years old, and one participant was aged 60 or above.

The interviews were conducted remotely via Zoom, following approval from our university’s Institutional Review Board (IRB). Participants were provided with an informed consent form and a demographic form when the schedule was fixed via email. At the beginning of each interview session, the interviewer introduced the study goals, interview process, and obtained participants’ verbal consent to proceed.

The interview protocol was designed to investigate the participants' experiences of using the GenAI tools and what kind of explanation formats contribute to more trustworthy GenAI tools. In the first part of the interview, we asked the participants to share their current interaction with the GenAI tools. The second part of the interview explored participants’ perceptions of trust and understanding of AI systems through two scenario-based discussions, examining how different explanation styles (e.g., descriptive details, rules, comparisons, and interactive dialogue) influenced accessibility, trust, and decision-making. These two scenarios include object identification and decision support. In the final part, participants were invited to share recommendations for developers and researchers on designing accessible explanation mechanisms that help BLV users better understand and trust AI response.

The interviews typically lasted between 45 and 60 minutes. All interviews were video-recorded with participants’ permission and later transcribed for analysis. Participants received a US\$30 Amazon gift card as compensation for their participation.

For data analysis, two authors independently participated in the coding process using a reflexive thematic analysis approach~\cite{braun2006using}. Inter-rater reliability was not calculated, as such metrics are not considered suitable for interview-based data where codes are not fixed~\cite{mcdonald2019reliability, o2020intercoder}. Instead, the authors met weekly to discuss codes, resolve differences to ensure consistency, and come up with themes.

\section{Findings}

Our analysis reveals that trust in AI for BLV users is not a static property of the system but a dynamic negotiation between the user's perception of risk and the system's mode of communication. We structure these findings across three dimensions: the calibration of trust relative to risk, the preference for conversational modalities to resolve ambiguity, and the strategies users employ to verify non-interpretable outputs.

\subsection{The Paradox of Trust: Risk Calibration and the Self-Blame Bias}
Participants demonstrated a sophisticated ability to calibrate trust based on the ``stakes'' of a given task. Trust was not binary; rather, it functioned as a sliding scale dependent on the consequence of error. P1 explicitly distinguished between low-stakes convenience and high-stakes health decisions:

\begin{quote}
``If it's medicine, I am not going to trust the AI 100\%. I will double-check. But if it is just a can of soda, I don't care if it is Coke or Pepsi.'' (P1)
\end{quote}

This distinction shows that users do not view AI reliability as a single fixed quality. Instead, they are willing to accept mistakes for convenient tasks while demanding strict accuracy for safety-critical situations. However, this calibration is frequently undermined by a pervasive ``Self-Blame Bias.'' When low-stakes errors occurred, participants often attributed the failure to their own actions, such as camera alignment or lighting, rather than questioning the model's capabilities. P3 described this internalization of failure as a primary barrier to effectively critiquing the system:

\begin{quote}
``It's garbage in, garbage out. If I don't give it a good picture... if I don't have the lighting right, or if I'm too close or too far away, it can't tell me what it is.'' (P3)
\end{quote}

The absence of specific feedback leads users to interpret model failure as a personal failure of execution. They assume the algorithm is functioning correctly, but it is simply limited by the quality of the input they provided. This tendency to assume ``user error'' protects the model from scrutiny. Unlike sighted users who can instantly verify if a model has hallucinated, BLV users often lack the ground truth to distinguish between a bad photo and a bad prediction. P2 highlighted how hallucinations shatter this trust entirely, as the user cannot easily verify which part of the description is real:

\begin{quote}
``Sometimes it hallucinates... it will make up things that aren't there just to give me an answer. That makes me really hesitant.'' (P2)
\end{quote}

\subsection{Beyond Description: Conversational Agents as Tools for Contestability}
A recurring frustration among participants was the ``Whole Scene Bias,'' where vision-to-language models prioritize generic scene descriptions over the user's specific intent. P4 expressed that standard captioning often obscures critical information with irrelevant background details:

\begin{quote}
``It's describing the whole scene, not a specific item. And what if you just want a specific item? ... I don't need to know there's a table there.'' (P4)
\end{quote}

These overly detailed descriptions create a significant problem with noise. By treating every visual detail as equally important, the model forces the user to mentally filter out irrelevant information to find the one specific detail they actually need. To mitigate this, participants overwhelmingly preferred conversational, ``back-and-forth'' modalities. This preference was not merely social; conversation served as a mechanism for functional contestability. It allowed users to refine the system's focus and correct initial misinterpretations. P3 noted that this dialogic approach transformed the interaction from a passive reception of information to an active negotiation:

\begin{quote}
``I like the back and forth... almost like a conversation. `Are you sure that's a red shirt?' and it says, `Oh, sorry, looking closer, it is actually maroon.''' (P3)
\end{quote}

This interaction shifts the control from the system back to the user. Rather than passively accepting a static text, the user can actively question the AI’s confidence and guide its attention toward the specific object they are looking for. Furthermore, P2 argued for a ``progressive disclosure'' architecture, where the system provides a high-level summary first, allowing the user to query for details only if necessary. \textit{``I would like a short description first... and then maybe a button or an option to say `tell me more'... I don't want a paragraph every time.''} shared P2. This approach reduces cognitive load during real-time navigation.

\subsection{Verification in the Dark: Proxy Strategies and the Human Fallback}
Lacking direct visual access to verify AI outputs, participants developed complex ``proxy'' strategies to triangulate the truth. These strategies often relied on redundancy or determinist data sources (like barcodes) to audit the probabilistic outputs of computer vision. P5 described a multi-modal verification process:

\begin{quote}
``I scan the barcode first. If the barcode says `Beans', then I see if the AI describes the can as beans. That's how I check.'' (P5)
\end{quote}

In this workflow, the barcode acts as a trusted fact that anchors the description. The user leverages this hard data to validate the often uncertain output of the computer vision model and builds trust through consistency rather than blind faith. Similarly, P1 utilized a ``multi-shot'' strategy, manually creating an ensemble of predictions to reach a confidence threshold that the system failed to provide:

\begin{quote}
``I usually take 3 or 4 pictures from different angles just to be sure. If it says `bottle' three times, then I believe it.'' (P1)
\end{quote}

The user is effectively performing the manual work of error correction that the system should handle automatically. While this strategy reduces the risk of a single bad prediction, it takes a significant amount of time and effort to perform for every object. However, these workarounds have an upper limit. When the AI's output remained ambiguous or inconsistent in high-stakes scenarios (e.g., reading currency or documents), participants hit a hard limit on interpretability. In these instances, P2 noted that the only viable solution was to abandon the automated system entirely for human assistance:

\begin{quote}
``If I really need to know, like for a document or money, and the AI is acting weird, I just hang up and call [a human volunteer].'' (P2)
\end{quote}

This reliance on human fallback indicates that current explainability methods fail to provide sufficient standalone evidence for users to resolve uncertainty without external aid.

\section{Discussion}
The central challenge for XAI in the non-visual context shifts from understanding model architecture to verifying immediate outcomes without visual ground truth. Our analysis reveals that the prevailing self-blame bias where users attribute system failures to their own photography rather than model uncertainty obscures the need for algorithmic accountability. To mitigate this, XAI design must transition from static descriptions to blame-aware interfaces that explicitly signal input quality issues and support functional contestability through conversational modalities. Our participants clearly preferred a ``back-and-forth'' interaction where they could ask to challenge the AI's confidence and force it to double-check its work. By allowing users to challenge the system's logic and demanding non-visual evidence rather than accepting a black-box output, future systems can bridge the verification gap that currently forces users to abandon automated tools for human assistance in high-stakes scenarios. This gap becomes more consequential in agentic settings. A user who misattributes an agent's error to their own input in one step of a multi-step task will continue participating in a pipeline that is already off course. The human fallback strategy our participants rely on is also no longer viable when the agent has already sent a message, submitted a form, or confirmed a transaction. Future agentic systems must support step-level attribution, distinguishing input quality failures from model errors at each stage, and must seek explicit user confirmation before taking irreversible actions. The conversational, back-and-forth modality our participants preferred maps directly onto this need: an agent that can say ``I selected this result because it matched your prior preference'' and accept a follow-up challenge is far safer for BLV users than one that acts silently.

\section{Conclusion}
This paper highlights how BLV users experience and evaluate the explainability in AI systems. Our findings demonstrate that trust is shaped by task risk and that users often internalize AI failures as their own mistakes. Participants consistently valued conversational explanations that allowed them to ask follow-up questions and resolve uncertainty. These results point to the limits of visually grounded XAI and emphasize the need for non visual, interactive explanation design. As AI systems take on more agentic roles, the cost of explanatory inaccessibility rises sharply. Multi-step pipelines, tool invocations, and irreversible actions make the self-blame bias more dangerous, the conversational verification strategies our participants valued more necessary, and the human fallback less available. Accessible agentic explainability is not a niche problem. It is a test of whether human-centered XAI can deliver on its core commitments.


\bibliographystyle{ACM-Reference-Format}
\bibliography{sample-base}

\appendix

\end{document}